\pdfoutput=1
\documentclass[]{article}
\usepackage{amsmath}
\usepackage{amssymb}
\usepackage{amsfonts}
\usepackage{authblk}
\usepackage{subfigure}
\usepackage{ctable}
\usepackage{multirow}
\usepackage{color}
\usepackage{booktabs}

\newcommand{\vecb}[1]{\mathbf{#1}}
\newcommand{\matb}[1]{\mathbf{#1}}
\newcommand{\R}{\mathbb{R}}

\begin{document}
\markboth{M. Romaszewski, P. G{\l}omb, P. Gawron}
{Natural hand gestures for human identification in a Human-Computer Interface}

\title{Natural hand gestures for human identification in a Human-Computer Interface}

\author{Micha{\l} Romaszewski}
\author{Przemys{\l}aw G{\l}omb}
\author{Piotr Gawron}
\affil{
Institute of Theoretical and Applied Informatics,Polish Academy of Sciences,
Ba\l{}tycka 5
\\
Gliwice, 44-100, Poland,
\\
\{michal,przemg,gawron\}@iitis.pl
}
\maketitle

\begin{abstract}
The goal of this work is the identification of humans based on motion data in
the form of natural hand gestures. In this paper, the identification problem is
formulated as classification with classes corresponding to persons' identities,
based on recorded signals of performed gestures. The identification performance
is examined with a database of twenty-two natural hand gestures recorded with
two types of hardware and three state-of-art classifiers: Linear Discrimination
Analysis (LDA), Support Vector machines (SVM) and k-Nearest Neighbour (k-NN).
Results show that natural hand gestures allow for an effective human
classification. 
\end{abstract}

\section{Introduction}
With a~widespread use of simple motion-tracking devices e.g. Nintendo Wii
Remote\texttrademark or accelerometer units in cell phones, the importance of
motion-based interfaces in Human-Computer Interaction (HCI) systems has become
unquestionable. Commercial success of early motion-capture devices led to the
development of more robust and versatile acquisition systems, both mechanical,
e.g. Cyberglove Systems Cyberglove\texttrademark,  Measurand
ShapeWrap\texttrademark, DGTech DG5VHand\texttrademark{} and optical  e.g.
Microsoft Kinect\texttrademark, Asus WAVI Xtion\texttrademark.  Also, the
interest in the analysis of a human motion itself 
\cite{McNeill:1992}, \cite{Quek:2002}, \cite{BergmannKopp:2010} 
has increased in the past few years.

While problems related to gesture recognition received much attention, an
interesting yet less explored problem is the task of recognising a human based
on his gestures. This problem has two main applications: the first one is the
creation of a gesture-based biometric authentication system, able to verify
access for authenticated users. The other task is related to personalisation of
applications with a motion component. In such scenario an effective classifier
is required to recognise between known users.

The goal of our experiment is to classify humans based on motion data in the
form of natural hand  gestures. Today's simple motion-based interfaces usually
limit users' options to a subset of artificial, well distinguishable gestures or
just detection of the presence of body motion. We argue that an interface should
be perceived by the users as natural and adapt to their needs. While modern motion-capture systems provide accurate recordings of human
body movement, creation of a HCI interface based on acquired data is not a
trivial task. Many popular gestures are ambiguous thus the meaning of a gesture
is usually not obvious for an observer and requires parsing of a complex
context. There are differences in body movement during the execution of a
particular gesture performed by different subjects or even in subsequent
repetitions by the same person. Some gestures may become unrecognisable with
respect to a particular capturing device, when important motion components are
unregistered, due to device limitations or its suboptimal calibration. We aim to
answer the question if high-dimensional hand motion data is distinctive enough
to provide a basis for personalisation component in a system with motion-based
interface.

In our works we concentrated on hand gestures, captured with two mechanical
motion-capture systems. Such approach allows to experiment with reliable
multi-source data, obtained directly from the device, without additional
processing. We used a gesture database of twenty two natural gestures performed
by a number of participants with varying execution speeds. The gesture database
is described in \cite{Glomb:2011}. We compare the effectiveness of three
established classifiers namely Linear Discrimination Analysis (LDA), Support
Vector machines (SVM) and k-Nearest Neighbour (k-NN). 

The following experiment scenarios are considered in this paper:
\begin{itemize}
\item Human recognition based on the performance of one selected gesture (e.g. 
`waving a hand',`grasping an object'). User must perform one specified gesture 
to be identified.
\item The scenario when instead of one selected gesture, a set of multiple 
gestures is used both for training and for testing. User must perform one of 
several gestures to be identified.
\item The scenario when different gestures are used for training and for 
testing of the classifier. User is identified based on one of several gestures, 
none of which were used for training the classifier.
\end{itemize}

The paper is organized as follows: Section 2 (Related work) presents a selection
of works on similar subjects, Section 3 (Method) describes the experiment,
results are presented in Section 4 (Results), along with authors' remarks on the
subject.

\section{Related work}
Existing approaches  to the creation of an HCI Interface that are based on
dynamic hand gestures can be categorized according to: the motion data gathering
method, feature selection, the pattern classification technique and the
domain of application.

Hand data gathering techniques can be divided into: device-based, where
mechanical or optical sensors are attached to a glove, allowing for measurement
of finger flex, hand position and acceleration, e.g.  \cite{Zhang:2009}, and
vision-based, when hands are tracked based on the data from optical sensors e.g.
\cite{Lahamy:2010}. A survey of glove-based systems for motion data
gathering, as well as their applications can be found in 
\cite{Dipietro:2008}, while \cite{Berman:2011} provides a comprehensive
analysis of the integration of various sensors into gesture recognition systems.

While non-invasive vision-based methods for gathering hand movement data are
popular, device-based techniques receive attention due to widespread use of
motion sensors in mobile devices. For example \cite{WandAccel:2006} presents
a high performance, two-stage recognition algorithm for acceleration signals,
that was adapted in Samsung cell phones.

Extracted features may describe not only the motion of hands but also their
estimated pose. A review of literature regarding hand pose estimation is
provided in \cite{Erol:2007}. Creation of a gesture model can be performed
using multiple approaches including Hidden Markov Models e.g. 
\cite{Mitra:2007} or Dynamic Bayesian Networks e.g. \cite{Xu:2006}. For
hand gesture recognition, application domains include: sign language recognition
e.g. \cite{Cooper_Visual:2011}, robotic and computer interaction e.g.
\cite{Lee:2006}, computer games e.g. \cite{Chan:2009} and virtual reality
applications e.g. \cite{Xu:2006}.

Relatively new application of HCI elements are biometric technologies aimed to
recognise a person based on their physiological or behavioural characteristic. A
survey of behavioural biometrics is provided in \cite{Yampolskiy:2008} where
authors examine types of features used to describe human behaviour as well as
compare accuracy rates for verification of users using different behavioural
biometric approaches. Simple gesture recognition may be applied for
authentication on mobile devices e.g. in \cite{Liu:2009} authors present a
study of light-weight user authentication system using an accelerometer while a
multi-touch gesture-based authentication system is presented in
\cite{Sae-Bae:2012}. Typically however, instead of hand motion more reliable
features like  hand layout \cite{Adan:2008} or body gait \cite{2003:Kale}
are employed. 

Despite their limitations, linear classifiers \cite{Koronacki:2005} proved to
produce good results for many applications, including face recognition
\cite{Wang:2004} and speech detection \cite{Martin:2001}. In
\cite{Rasamimanana:2006} LDA is used for the estimation of consistent
parameters to three model standard types of violin bow strokes. Authors show
that such gestures can be effectively presented in the bi-dimensional space.  In
\cite{Moiz:2011}, the LDA classifier was compared with neural networks (NN)
and focused time delay neural networks (TDNN) for gesture recognition based on
data from a 3-axis accelerometer. LDA gave similar results to the NN approach,
and the TDNN technique, though computationally more complex, achieved better
performance. An analysis of LDA and the PCA algorithm, with a discussion about
their performance for the purpose of object recognition is provided in
\cite{Martinez:2001}. SVM and k-NN classifiers were used in
\cite{Zhang:2006} for the purpose of visual category recognition. A
comparison of the effectiveness of these method is classification of human gait
patterns is provided in \cite{Sudha:2012}.

Thorough analysis of a gesture dataset used in the experiments, along with a
discussion on the benefits of naturality of HCI interface elements can be found
in \cite{Glomb:2012}. PCA analysis of the same dataset together with
visualization of eigengestures can be found in \cite{GawronEtAll:2011}.

\section{User identification using classification of natural gestures}
The general idea is to recognise a gesture performer. Experiment data consist of
data from `IITiS Gesture Database' that contains natural gestures performed by
multiple participants. Three classifiers will be used. PCA will be performed on
the data to reduce its dimensionality.

\subsection{Experiment data}

\begin{figure}[h]
 \centering
\subfigure[]{\includegraphics[scale=1.0]{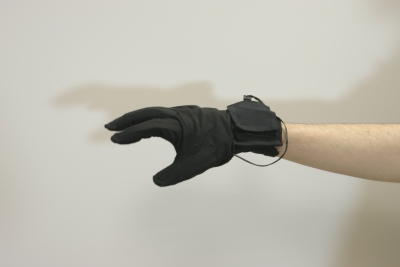}}
\subfigure[]{\includegraphics[scale=1.0]{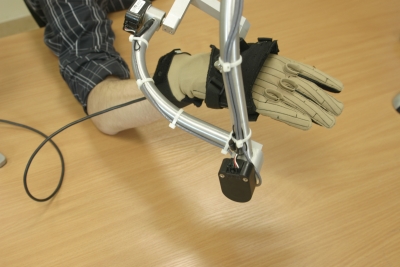}}
  \caption{Scenes from recording of `IITiS Gesture Database'. Left DG5VHand glove, right CyberGlove/CyberForce system.}
  \label{fig:gestures_intro}
\end{figure}

A set of twenty-two natural hand gesture classes from `IITiS Gesture
Database'\footnote{The database can be downloaded from
http://gestures.iitis.pl/} \cite{Glomb:2011},
Tab.~\ref{id:table:gestureList}, was used in the experiments. Gestures used in
this experiments were recorded with two types of hardware (see Fig.
\ref{fig:gestures_intro}). First one was the DGTech 
DG5VHand\texttrademark{}\footnote{http://www.dg-tech.it/vhand} motion capture 
glove \cite{DG5VHand}, containing 5 finger bend sensors (resistance type),
and  three-axis accelerometer producing three acceleration and two orientation 
readings. Sampling frequency was approximately 33 Hz. The second one was
Cyberglove Systems CyberGlove\texttrademark{} 
\footnote{http://www.cyberglovesystems.com/products/cyberglove-ii/overview} 
with a CyberForce\texttrademark{} System for position and  orientation
measurement. The device produces 15 finger bend, three position  and four
orientation readings with a frequency of approximately 90 Hz.

\newcounter{gestureCounter}\stepcounter{gestureCounter}
\newcommand{\gc}{\arabic{gestureCounter}\stepcounter{gestureCounter}}
\ctable[
  star,
  cap = Gesture list,
  caption = The gesture list used in experiments. Source: 
  \cite{Glomb:2011} , 
  label = id:table:gestureList,
]{cllll}{
  \tnote{We use the terms `symbolic', `deictic', and `iconic' based on McNeill 
\& Levy \cite{McNeill:1992} classification, supplemented with a category of 
`manipulative' gestures (following \cite{Quek:2002})}
  \tnote[b]{Significant motion components: T-hand translation, R-hand rotation, 
F-individual finger movement}
  \tnote[c]{This gesture is usually accompanied with a specific object 
(deictic) reference}
}{\FL
& Name & Class\tmark & Motion\tmark[b] & Comments\ML
\gc & \emph{A-OK} & symbolic & F & common `okay' gesture\NN
\gc & \emph{Walking} & iconic & TF & fingers depict a walking person\NN
\gc & \emph{Cutting} & iconic & F & fingers portrait cutting a sheet of paper\NN
\gc & \emph{Showe away} & iconic & T & hand shoves avay imaginary object\NN
\gc & \emph{Point at self} & deictic & RF & finger points at the user\NN
\gc & \emph{Thumbs up} & symbolic & RF & classic `thumbs up' gesture\NN
\gc & \emph{Crazy} & symbolic & TRF & symbolizes `a crazy person'\NN
\gc & \emph{Knocking} & iconic & RF & finger in knocking motion\NN
\gc & \emph{Cutthroat} & symbolic & TR & common taunting gesture\NN
\gc & \emph{Money} & symbolic & F & popular `money' sign\NN
\gc & \emph{Thumbs down} & symbolic & RF & classic `thumbs down' gesture\NN
\gc & \emph{Doubting} & symbolic & F & popular flippant `I doubt'\NN
\gc & \emph{Continue} & iconic\tmark[c] & R & circular hand motion `continue', 
`go on'\NN
\gc & \emph{Speaking} & iconic & F & hand portraits a speaking mouth\NN
\gc & \emph{Hello} & symbolic\tmark[c] & R & greeting gesture, waving hand 
motion\NN
\gc & \emph{Grasping} & manipulative & TF & grasping an object\NN
\gc & \emph{Scaling} & manipulative & F & finger movement depicts size change\NN
\gc & \emph{Rotating} & manipulative & R & hand rotation depicts object 
rotation\NN
\gc & \emph{Come here} & symbolic\tmark[c] & F & fingers waving; `come here'\NN
\gc & \emph{Telephone} & symbolic & TRF & popular `phone' depiction\NN
\gc & \emph{Go away} & symbolic\tmark[c] & F & fingers waving; `go away'\NN
\gc & \emph{Relocate} & deictic & TF & `put that there'\LL
}

During the experiment, each participant was sitting at the table with the motion
capture glove on their right hand. Before the start of the experiment, the hand
of the participant was placed on the table in a fixed initial position. At the
command given by the operator sitting in front of the participant, the
participant performed the gestures. Each gesture was performed six times at
natural pace, two times at a rapid pace and two times at a slow pace. Gestures
number 2, 3, 7, 8, 10, 12, 13, 14, 15, 17, 18, 19, 21 are periodical and in
their case a single performance consisted of three periods. The termination of
data acquisition process was decided by the operator.

\subsection{Dataset exploration}

\begin{figure}[h]
 \centering
\subfigure[DG5VHand\texttrademark]{\includegraphics[scale=1]{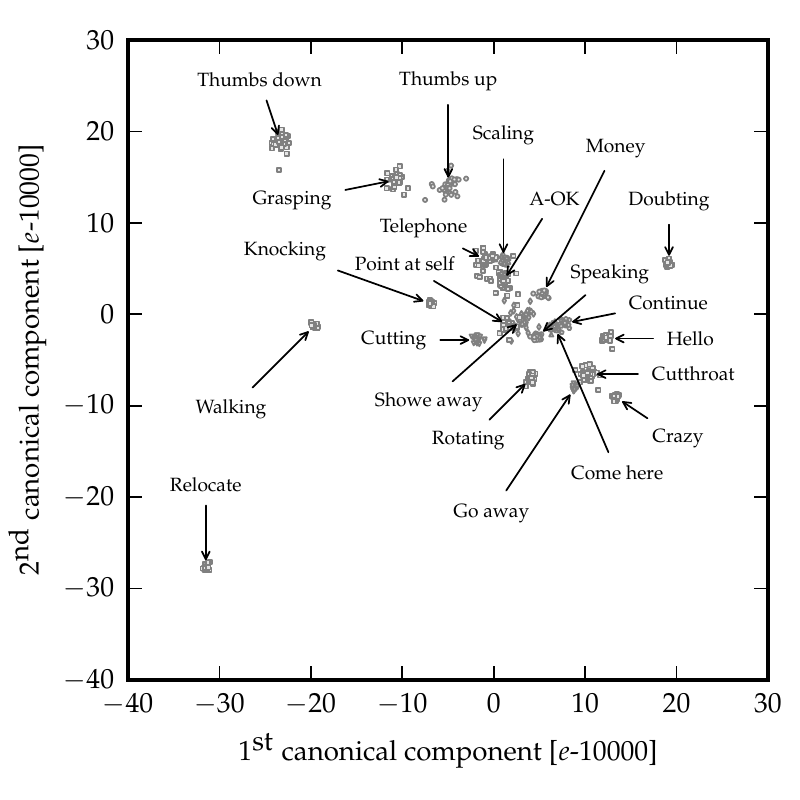}}
\subfigure[CyberGlove\texttrademark ]{\includegraphics[scale=1]{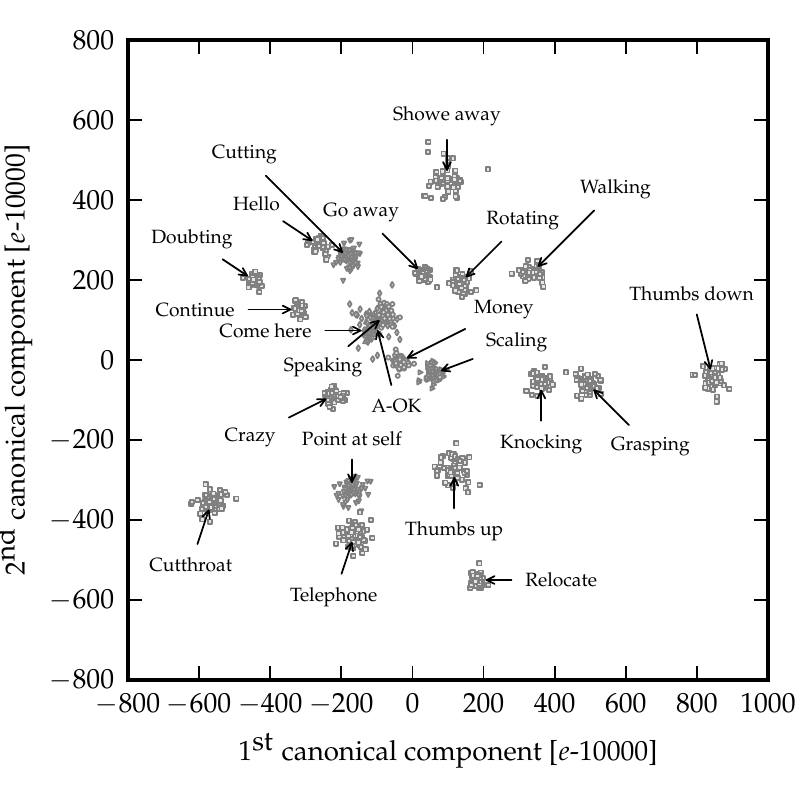}}
  \caption{Visualisation of data separability gestures dataset with use of LDA. The original data is projected on $d=2$ canonical vectors.}
  \label{fig:lda_projection}
\end{figure}

Figure \ref{fig:lda_projection} presents the result of performing LDA (further described in subsection \ref{LDA}) on the
dataset: projection of the dataset on the first two components of 
$\mathbf{W}^{-1}\mathbf{B}$ for both devices. It can be observed that many
gestures are linearly separable.  In the majority of visible gesture classes,
elements are centred around their respectable mean, with an almost uniform
variance. Potential conflicts for small number of gestures may be observed for
local regions of the projected data space.

\subsection{Data preprocessing}
\label{id:section:preprocessing}
A motion capture recording performed with a~device with $m$ sensors
generates a time sequence of vectors $\vecb{x}_{t_i}\in \mathbb{R}^m$. For the 
purpose of our work each recording was linearly interpolated and re-sampled to 
$t=100$ samples, generating data matrices
$\matb{A}_l=[x_{l}^{(ij)}]\in\R^{m\times t}$, where $l$ enumerates  recordings.
Then data matrices were normalized by computing the
t-statistics 
\begin{equation*}
\matb{A}'_l=\frac{x_{l}^{(ij)}-\bar{x}_i}{\sigma_i}, 
\end{equation*}
where $\bar{x}_i$, $\sigma_i$ are mean and standard deviation for a given 
sensor $i$ taken over all $l$ recordings in the database.

Subsequently every matrix $\matb{A}'_l$ for was vectorized row-by-row, so that
it was transformed into data vector 
\begin{equation*}
\vecb{x}_l=[x^{(11)}_l,\ldots,x^{(m1)}_l,\ldots,x^{(1t)}_l,\ldots,x^{(mt)}_l]^T,
\end{equation*}
belonging to $\R^p$, $p=m t$. 
These data vectors were organised into $n=4$ (for DG5VHand) and $n=6$ (for
Cyberglove) classes $C_k$ corresponding to participants registered
with each device.
\subsection{PCA}
Principal Component Analysis \cite{Wall:2003} may be defined as follows. 
Let 
$\mathbf{X}=[\mathbf{x}_1,\mathbf{x}_2\ldots,\mathbf{x}_L]$
be the data matrix, where $\mathbf{x}_i\in\R^{p}$ are data vectors with zero
empirical mean. The associated covariance matrix is given by
$\mathbf{\Sigma}=\mathbf{X}\mathbf{X}^T$. 
By performing eigenvalue decomposition of 
$\mathbf{\Sigma}=\mathbf{O}\mathbf{\Lambda}\mathbf{O}^T$ such that 
eigenvalues  $\lambda_i, i=1,..,p$ of $\mathbf{\Lambda}$ are ordered in 
descending order
$\lambda_1\geq\lambda_2\geq\ldots\geq\lambda_p>0,$
one obtains the sequence of principal components 
$[\mathbf{o}_{1},\mathbf{o}_{2},\ldots,\mathbf{o}_{p}]$ which are columns of 
$\mathbf{O}$ \cite{Wall:2003}.
One can form a feature vector $\mathbf{y}$ of dimension $p'\leq p$ by 
calculating 
$\mathbf{y} = [\mathbf{o}_{1},\mathbf{o}_{2},\ldots,\mathbf{o}_{p'}]^T \mathbf{x}.$
\subsection{LDA} \label{LDA}
Linear Discriminant Analysis--thoroughly presented in
\cite{Koronacki:2005}--is a supervised, discriminative technique producing an
optimal linear classification function, which transforms the data from
$p'$ dimensional space
$\mathbb{R}^{p'}$ into a lower-dimensional classification space $\R^d$.

Let the \emph{between-class scatter matrix} $\matb{B}$ be defined as follows
\begin{equation*}
\mathbf{B}=\frac{1}{k-1}
\sum_{i=1}^k 
n_i(\bar{\vecb{x}}_i-\bar{\vecb{x}})(\bar{\vecb{x}}_i-\bar{\vecb{x}})^T,
\end{equation*}
where $\bar{\vecb{x}}$ denotes mean of class means $\bar{\vecb{x}}_i$ i.e.	
$\bar{\vecb{x}}=\frac{1}{k}\sum_{i=1}^k \bar{\vecb{x}}_i$, and $n_i$ is the 
number of samples in class $i$. 
Let \emph{within-class scatter matrix} $\mathbf{W}$ be
\begin{equation*}
\mathbf{W}=\frac{1}{n-k}\sum_{j=1}^k \sum_{\vecb{x}_i\in C_j} 
(\vecb{x}_i-\bar{\vecb{x}}_j)(\vecb{x}_i-\bar{\vecb{x}}_j)^T,
\end{equation*}
where $n$ is the total number of the samples in all classes.

The eigenvectors of matrix $\matb{W}^{-1}\matb{B}$ ordered by their respective
eigenvalues are called the canonical vectors. By selecting first $d$ canonical
vectors and arranging them row by row as the projection matrix 
$\tilde{\mathbf{A}}^{(d)}\in
\R^{d\times p'}$ any vector $\vecb{x}\in \R^{p'}$ can be projected onto a
lower-dimensional feature space ${\mathbb{R}^d}$. Using LDA one can effectively 
apply simple classifier e.g. for $k$-class problem. A vector $\vecb{x}$ is 
classified to class $C_j$ if following inequality is observed 
$
||\tilde{\matb{A}}^{(d)} (\vecb{x} - \bar{\vecb{x}}_j)||<
||\tilde{\matb{A}}^{(d)} (\vecb{x} - \bar{\vecb{x}}_k)||,
\label{id:equation:classification_mc}
$for all $k\neq j$. $||\cdot||$ denotes Euclidean norm.

Note that when the amount of available data is limited, LDA technique may result
in the matrix $\mathbf{W}$ that is singular.
In this case one can use Moore-Penrose pseudoinverse \cite{Tian:88}.
Matrix $\matb{W}^{-1}$ is replaced by Moore-Penrose pseudoinverse matrix
$\matb{W}^{\dagger}$ and canonical vectors are eigenvectors of the matrix
$\matb{W}^{\dagger}\matb{B}$.

\subsection{$k$-NN} \label{KNN}
The $k$-Nearest Neighbour ($k$-NN) method \cite{Hechenbichler:2004}
classifies the sample by assigning it to the  most frequently represented class
among k nearest samples. It may be described as follows. Let 
\begin{equation*}
L=\{(y_i,\mathbf{x}_i),i=1,..,n_L\}
\end{equation*}
be a training set where $y_i\in\{1,..,c\}$ denotes class labels, and
$\mathbf{x}_i\in\R^{p}$, are feature vectors. For a nearest neighbour 
classification, given a new observation $\mathbf{x}$, first a nearest element 
$(y_{i_1},\mathbf{x}_{i_1})$ of a learning set is determined 
\begin{equation*}
i_1 = \mathop{\mathrm{argmin}}_{i}(d(\mathbf{x},\mathbf{x}_i))
\end{equation*}
with Euclidean distance $d(\cdot,\cdot)$
\begin{equation*}
d(\mathbf{x},\mathbf{x}_i)=\sqrt{(\mathbf{x} - \mathbf{x}_i)^\top(\mathbf{x} - \mathbf{x}_i)}
\end{equation*}
and resulting class label is $y_{i_1}$.

Usually, instead of only one observation from $L$, $k$ most similar elements are
considered. Therefore, counts of class labels for 
$Y=\{y_{i_1}, \dots, y_{i_k}\}$ are determined for each class
\begin{equation*}
K_i=\sum_{y\in Y} \delta_{iy}
\end{equation*}
where $\delta_{iy}$ denotes Dirac delta.
The class label is determined as most common class present in the results
\begin{equation*}
y = \mathop{\mathrm{argmax}}_i \{K_1,\dots,K_c\}.
\end{equation*}

Note that in case of multiple classes or single class and even $k$ there may be
a tie in the top class counts; in that case results may be dependent on data
order and behaviour of $\mathrm{argmax}$ implementation.

\subsection{SVM} \label{SVM}
Support Vector machines (SVM) presented in \cite{Byun:2002} are supervised
learning methods based on the principle of constructing a hyperplane separating
classes with the largest margin of separation between them. The margin is the
sum of distances from the hyperplane to closest data points of each class. These
points are called Support Vectors. SVMs  can be described as follows. Let 
\begin{equation*}
L=\{(\mathbf{x}_i,y_i),i=1,..,n_L\}, \mathbf{x}_i\in\R^{p}
\end{equation*}
be a set of linearly separable training samples where $y_i\in\{-1,1\}$ denotes 
class labels. We assume the existence of a $p$-dimensional hyperplane ($\cdot$
denotes dot product)
\begin{equation*}
\mathbf{w}\cdot \mathbf{x}+b = 0,
\end{equation*}
 separating $\mathbf{x}$ in $\R^{p}$.

The distance between separating hyperplanes satisfying $|\mathbf{w}\cdot 
\mathbf{x}+b|=1$ and $|\mathbf{w}\cdot \mathbf{x}+b|=-1$ is 
$\frac{2}{||\mathbf{w}||}$. The optimal separating hyperplane can be found by 
minimising 
\begin{equation}
\label{equation:svm:2}
\min(\mathbf{w})=\frac{{||\mathbf{w}||}^2}{2}=\frac{\mathbf{w} \cdot 
\mathbf{w}}{2}, 
\end{equation}
under the constraint
\begin{equation}
\label{equation:svm:1}
y_i (\mathbf{x}_i\cdot \mathbf{w}+b) \geq 1.
\end{equation}
for all $\mathbf{x}_i, i=1,..,n_L$.

When the data is not linearly separable, a hyperplane that maximizes the margin
while minimizing a quantity proportional to the misclassification errors is
determined by introducing positive slack variables $\xi_i$ in the equation
\ref{equation:svm:1}, wchich becomes:
\begin{equation}
\label{equation:svm:3}
y_i (\mathbf{x}_i\cdot \mathbf{w}+b) \geq 1+\xi_i.
\end{equation}
and the equation (\ref{equation:svm:2}) is changed into:
\begin{equation}
\label{equation:svm:4}
\min(\mathbf{w})=\frac{\mathbf{w} \cdot \mathbf{w}}{2}+C\sum_{i=1}^{n}\xi_i, 
\end{equation}
where $C$ is a penalty factor chosen by the user, that controls the trade off 
between the margin width and the misclassification errors. 

When the decision function is not a linear function of the data, an initial
mapping $\phi$ of the data into a higher dimensional Euclidean space $H$ is
performed as $\phi:\R^{n_L} \rightarrow H$ and the linear classification problem
is formulated in the new space. The training algorithm then only depends on the
data through dot product in $H$ of the form $\phi(\mathbf{x}_i)\cdot
\phi(\mathbf{x_j})$.  The Mercer's theorem \cite{Burges:1998} allows to
replace $\phi(\mathbf{x}_i)\cdot \phi(\mathbf{x_j})$ by a positive definite
symmetric kernel function $K(\mathbf{x_i},\mathbf{x_j})$, e.g. Gaussian
radial-basis function $K(\mathbf{x_i},\mathbf{x_j}) =
\exp(-\gamma||\mathbf{x_i}-\mathbf{x_j}||^2)$, for $\gamma>0$.

\section{Results}
Our objective was to evaluate the performance of user identification based on
performed gestures. To this end, in our experiment class labels are assigned to
subsequent humans performing gestures (performers' ids were recorded during
database acquisition). Three experiment scenarios were investigated, differing
by the range of gestures used for recognition. The three  classification methods
described before were used, evaluated in two-stage $k$-fold cross validation
scheme.

\subsection{Scenarios}
Three scenarios related to data labelling were prepared:
\begin{itemize}
\item Scenario A. Human classification using specific gesture. Each gesture was
treated as a separated case, and a classifier was created and verified using
samples from this particular gesture.
\item Scenario B. Human classification using a set of known gestures. Data from
multiple gestures was used in the experiment. Whenever the data was divided into
a teaching and testing subset, proportional amount of samples for each gesture
were present in both sets.
\item Scenario C. Human classification using separate gesture sets. In this
scenario the data from multiple gestures was used, similarly to Scenario B.
However, teaching subset was created using different gestures than a testing
subset.
\end{itemize}

\subsection{Experiments}

The three classifiers were used, with the following parameter ranges:
\begin{itemize}
\item LDA, with number of features $n=3,5,10,15,20,25,30,35$;
\item $k$-NN, with number of neighbours $k=1,2,3,4,5,7,10,20,30,40,50$;
\item SVM, with Radial Basis Function (RBF) and 
$C,\gamma \in \langle 0.001, 1.0 \rangle$.
\end{itemize} 

Common parameters values found by cross-validation: 
\begin{itemize}
\item LDA, $3-5$ features
\item $k$-NN, $1-3$ neighbours for Scenarios B,C, $20-50$ for Scenario C;
\item SVM, 
$\gamma \in \langle 0.001, 0.005\rangle$ $C \in \langle 0.001, 0.01\rangle$.
\end{itemize} 

The parameter selection and classifier performance evaluation was performed by
splitting the available data into training and testing subset in two-stage
$k$-fold cross validation (c.v.) scheme, with $k=4$. Inner c.v. stage
corresponds to grid search parameter optimization and model selection. The outer
stage corresponds to final performance evaluation. The PCA was performed on the
whole data set before classifier training. The amount of principal components
was chosen empirically as $p'=100$.

\subsection{Results and discussion}

\begin{table}[h]
\centering
\begin{tabular}{ccccccc}
\toprule
Scenario&\multicolumn{6}{c}{Accuracy}\\ 
\cmidrule(r){2-7}
&\multicolumn{3}{c}{DG5VHand}&\multicolumn{3}{c}{CyberGlove}\\ 
\cmidrule(r){2-4}\cmidrule(r){5-7}
&LDA&k-NN&SVC&LDA&k-NN&SVC\\
\midrule
A&$97$&$94.7$&$96.2$&$99.4$&$99.7$&$99.9$\\
B&$88.9$&$94.7$&$94.8$&$99.6$&$99.7$&$100$\\
C&$75$&$52.3$&$73.8$&$92.8$&$69.3$&$89.9$\\
\bottomrule
\end{tabular} 
\caption{Classification accuracy (\%) for three considered scenarios.}
\label{id:table:mean_accuracy}
\end{table}

The accuracy of the classifiers for three discussed scenarios is presented in
Tab. \ref{id:table:mean_accuracy}. Confusion matrices for experiments B,C are
presented  on Figure \ref{fig:confusion_matrices}. High classification accuracy
can be observed for scenarios $A$ and $B$ when a classifier is provided with
training data for specific gesture. In scenario C, however, the accuracy
corresponds to a situation when a performer is recognised based on an unknown 
gesture. While the classification accuracy is lower than in previous scenarios,
it should be noted that the classifier was created using a limited amount of
high-dimensional data. The difference between the accuracy for both devices can
be explained by significantly higher precision of a CyberGlove device, where
hand position is captured using precise rig instead of an array of
accelerometers. 

\begin{table}[ht]
\centering
\begin{tabular}{c| c c c c}
& & LDA & $k$-NN & SVC\\ \multirow{2}{*}{B}
&{\begin{sideways}\parbox{5cm}{\centering DG5VHand}\end{sideways}}
&\includegraphics[scale=0.6]{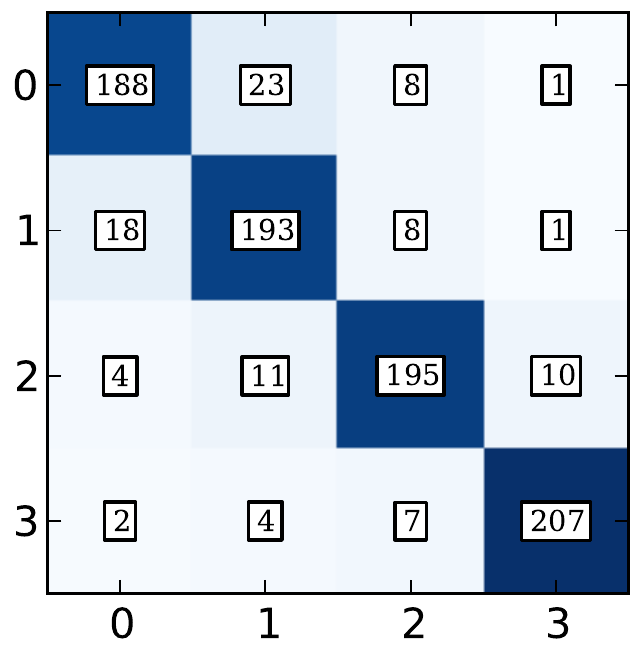}
&\includegraphics[scale=0.6]{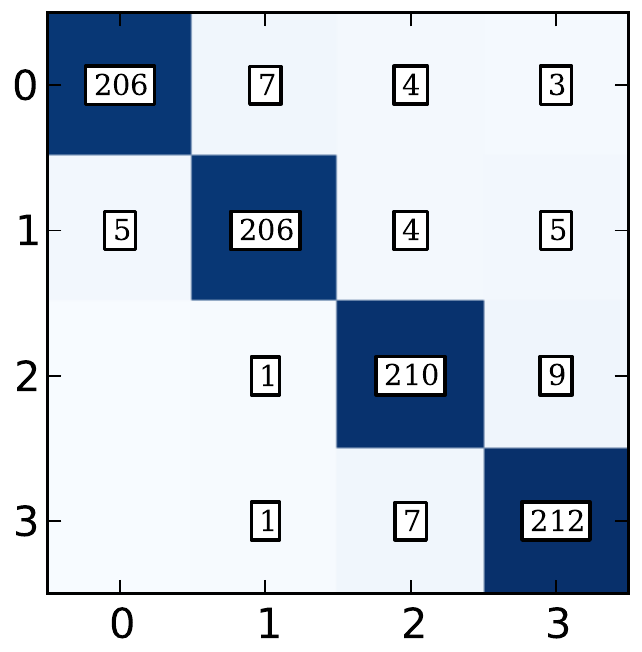}
&\includegraphics[scale=0.6]{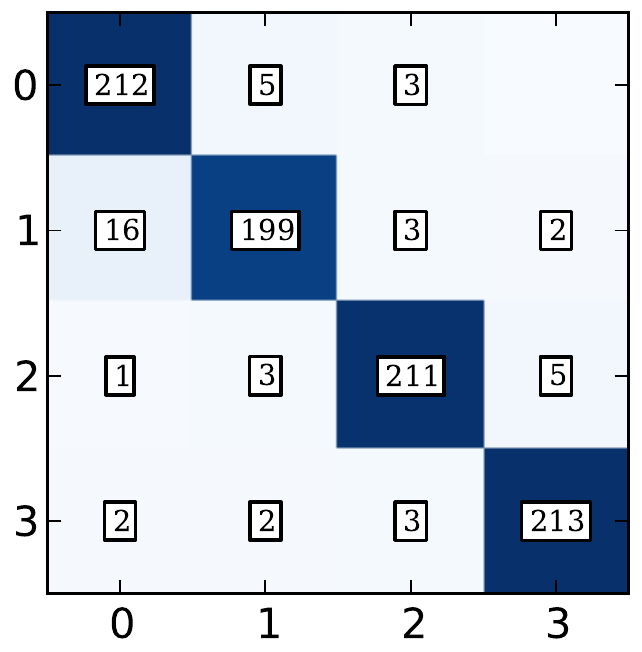}\\ 
&{\begin{sideways}\parbox{5cm}{\centering CyberGlove}\end{sideways}}
&\includegraphics[scale=0.6]{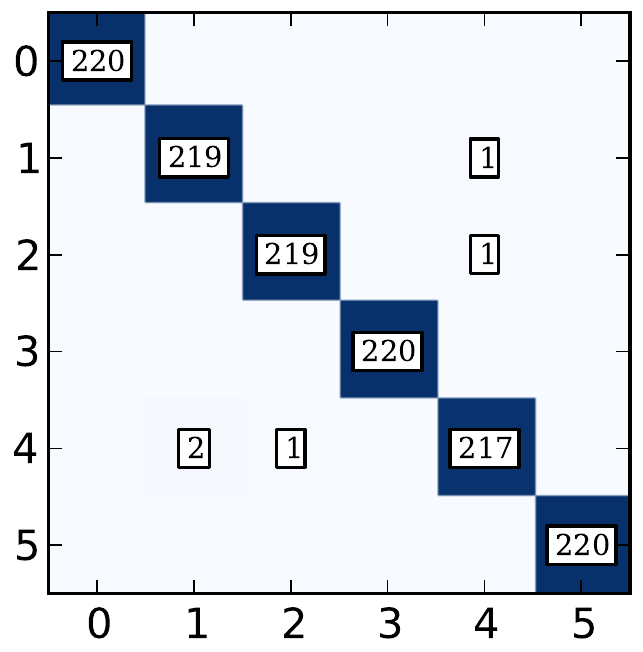}
&\includegraphics[scale=0.6]{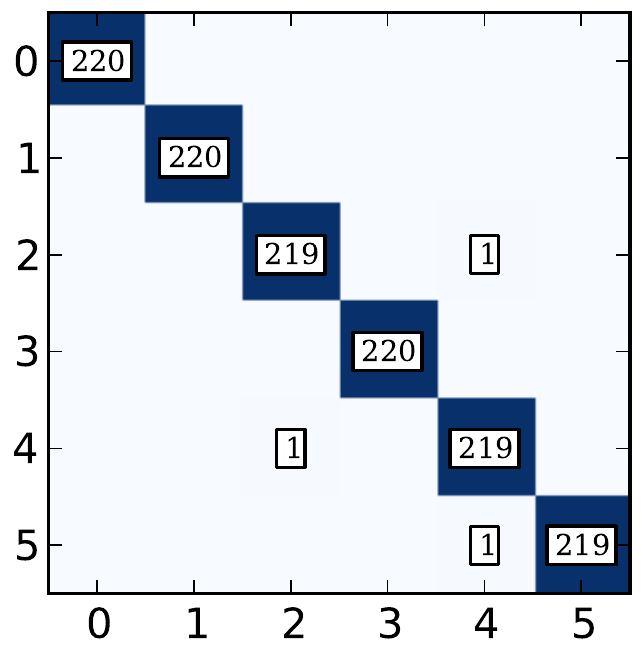}
&\includegraphics[scale=0.6]{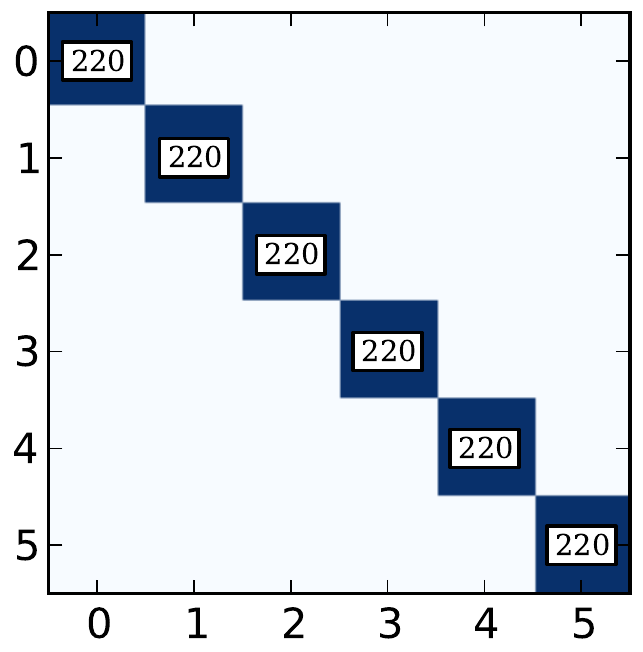}\\ 
\hline 
\multirow{2}{*}{C}
&{\begin{sideways}\parbox{5cm}{\centering DG5VHand}\end{sideways}}
&\includegraphics[scale=0.6]{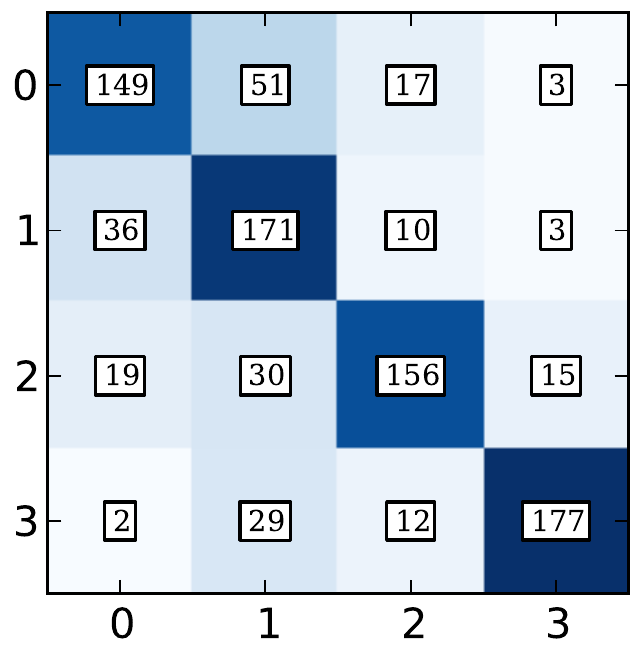}
&\includegraphics[scale=0.6]{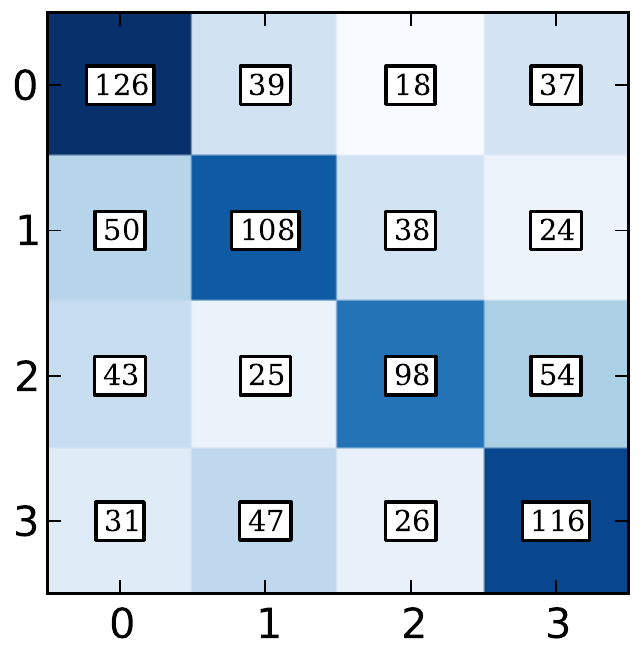}
&\includegraphics[scale=0.6]{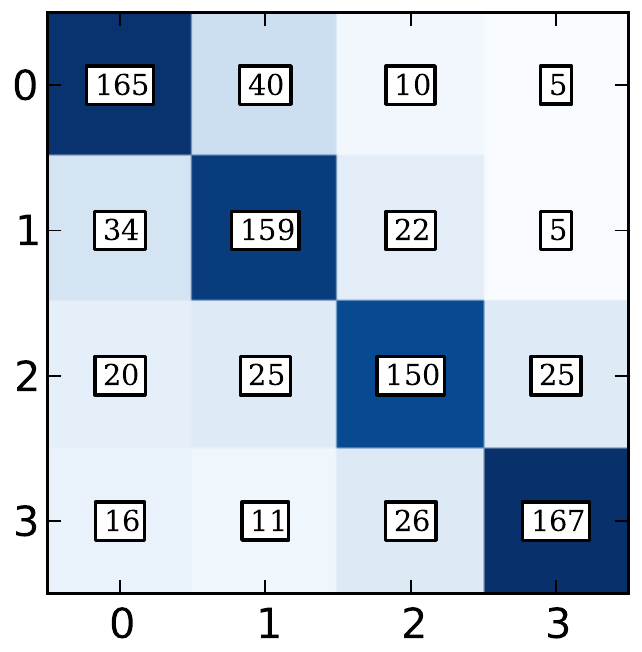}\\ 
&{\begin{sideways}\parbox{5cm}{\centering CyberGlove}\end{sideways}}
&\includegraphics[scale=0.6]{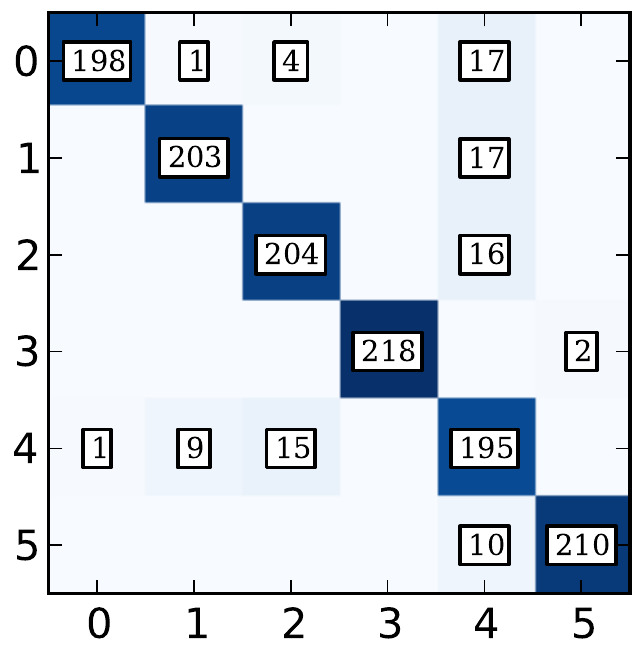}
&\includegraphics[scale=0.6]{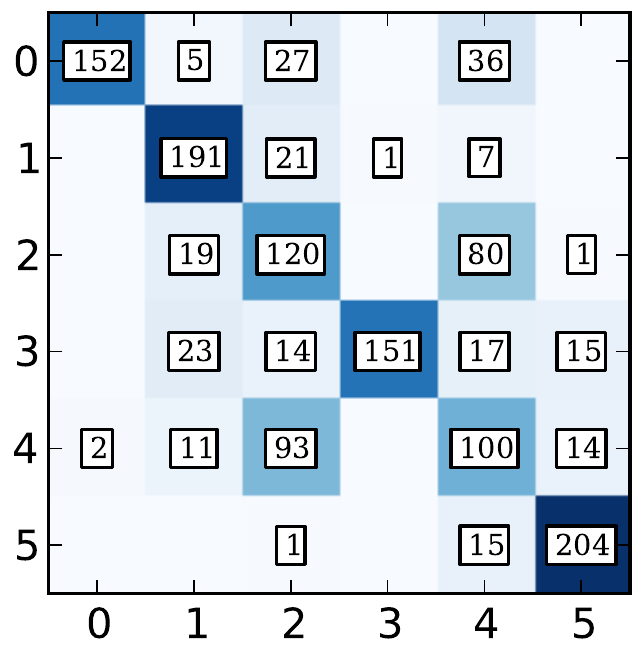}
&\includegraphics[scale=0.6]{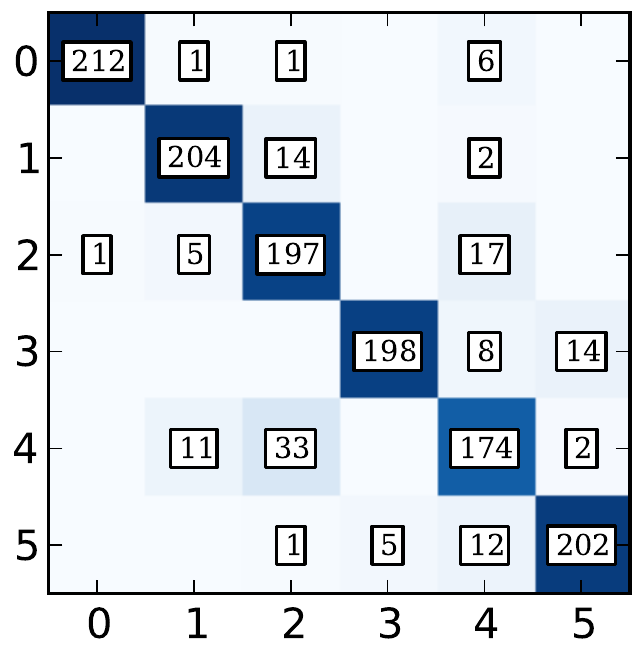} 
\end{tabular}
\caption{Confusion matrices for scenarios B and C, all classifiers and both devices.}
\label{fig:confusion_matrices}
\end{table} 

Results of the experiment show that even linear classifiers can be successfully
employed for recognition of human performers based on their natural gestures.
Relatively high accuracy for experiment C indicates that the general
characteristics of a human natural body movement is highly discriminative, even
for different gesture patterns. While mechanical devices used in experiments
provide accurate measurements of body movements, they may be replaced by less
cumbersome data gathering device e.g. Microsoft Kinect\texttrademark. 

\section{Conclusion}
Experiments confirm that natural hand gestures are highly discriminative and
allow for an accurate classification of their performers. Applications of such
solution allow e.g. to personalise tools and interfaces to suit the needs of
their individual users.  However, a separate problem lies in the detection of
particular gesture performer based on general hand motion. Such task requires
deeper understanding of motion characteristics as well as identification of
individual features of human motion.

\section*{Acknowledgements}
The work of M. Romaszewski and P. G{\l}omb has been partially supported  by the
Polish Ministry of Science and Higher Education project NN516482340
``Experimental station for integration and presentation of 3D views''. Work by
P. Gawron was partially suported by  Polish Ministry of Science and Higher
Education project NN516405137 ``User interface based on natural gestures  for
exploration of virtual 3D spaces''. Open source machine learning library
scikit-learn \cite{scikit-learn} was used in experiments. We would like to
thank Z.~Pucha{\l}a and J. Miszczak for fruitful discussions.

\end{document}